\documentclass[preprint,12pt]{elsarticle}

\usepackage{amssymb}

\usepackage{tabularx}
\usepackage{booktabs}
\usepackage{graphicx}

\begin{document}

\begin{frontmatter}



\title{A first-principle study of half-Heusler alloys CKMg and SiKMg.}


\author[mymainaddress]{Gang Lei}
\author[mymainaddress]{Xiao-Xiong Liu}
\author[mymainaddress]{Lei Li}
\author[mymainaddress]{Qiang Gao}
\author[mymainaddress]{Xian-Ru Hu}

\author[mymainaddress]{Jian-bo Deng\corref{mycorrespondingauthor}}
\cortext[mycorrespondingauthor]{Corresponding author}
\ead{Dengjb@lzu.edu.cn}

\address[mymainaddress]{School of Physical Science and Technology, Lanzhou University,
 Lanzhou 730000, People's Republic of China}

\begin{abstract}
The structural, electronic, and magnetic properties of half-Heusler alloys CKMg and SiKMg are studied by using first-principle density functional theory. The calculations reveal the SiKMg alloy is a half-metallic ferromagnet with the magnetic moment of 1 $\mu_B$ per formula unit at equilibrium lattice constant. The magnetic moment mainly originates from the strong spin-polarization of $p$ electrons of Si atom and partial involvement of $d$ electrons of K atom. The half-metallic gap is 0.105 eV.  The robustness of half-metallic against the lattice constants for SiKMg is also calculated. CKMg alloy is nearly half-metallic with a spin polarization of 99.99 \%\ at equilibrium lattice constant, but it is a good half-metallic alloy  when a low pressure is applied. This shows CKMg is a very promising spintronic functional material.
\end{abstract}

\begin{keyword}
First-principle \sep half-Heusler  alloy \sep Magnetic property \sep Pressure effect


\end{keyword}

\end{frontmatter}


\section{Introducion}
\label{}

Half-metallic ferromagnets (HMF), of which one of the two channels shows
metallic behaviour while the other shows the semiconductor or insulator
behaviour at the Fermi level, have been attracting more and more interests due
to their great potential applications in spintronic devices. Since the
conception of the half-metal (HM) was introduced by de Groot et al. \cite{1} in
1983, many HMs have been found in a series of materials, such as ferromagnetic
metallic oxides \cite{2,3,4,5}, Heusler compounds \cite{6,7,8,9,10,11,12}, binary transition metal
pnictides, chalcogenides with zinc-blended, rocksalt structures \cite{13,14,15,16,17}, etc.

Among these materials the half-metallic Heusler alloys are expected to play a
key role in practical applications as a result of their very high Curie
temperature and structural similarity compared to the widely used binary
semiconductors crystallizing in the zinc-blended structure. Many researches
have been made on these alloys. Based on first-principle electronic structure
calculations, lots of them are predicted to be HMF. However, the half-metallic
properties are \ difficult to obtain at room temperature because the
half-metallic gap of the full-Heusler compounds is not very large.
Furthermore, many half-metallic alloys which contain 3$d$ transition metals
usually have a big magnetic moment. They are not very suitable for spintronic
device application, because the large magnetic moment means high stray fields
and large energy losses. This defect motivates us to search for new HM
materials which do not contain 3$d$ transition metals.

Based on first-principle study, we investigate the stability in structure,
magnetic and pressure effect of  CKMg and SiKMg alloys. This kind of half-Heusler
compounds which has the general formula XYZ and crystallizes in
non-centrosymmetric cubic MgAgAs(C1$_b$)  structure with
FB${\overline{4}}3m$ (space group No.216), has three interpenetrating $fcc$ lattices.
The corresponding Wyckoff positions are $r_1$= (0.00, 0.00, 0.00), $r_2$=
(0.25, 0.25, 0.25), $r_3$= (0.75, 0.75, 0.75) \cite{18}, and each of the X, Y and Z
atoms can occupy one of the three positions. Then interchanging the positions
of atoms in the cubic structure, we can get three phases, which are $\alpha$=
($r_1$, $r_2$, $r_3$) phase, $\beta$= ($r_3$, $r_1$, $r_2$) phase and
$\gamma$= ($r_2$, $r_3$, $r_1$) phase, respectively.

\section{Computational method}
\label{}
In this paper, we perform our calculations by using full-potential
local-orbital minimum-basis code (FPLO) \cite{19, 20}. The scalar relativistic
exchange-correlation energy is treated within the generalized gradient
approximation (GGA) \cite{21}. The $k$ mesh is performed as 24. Accurate Brillouin zone
integrations are performed by using the standard special $K$ point technique
of the tetrahedron. The convergence cretia of a self-consistent field
iteration is set to both the density ($10^{- 6}$ in code specific units) and the total
energy ($10^{- 8}$ hartree).

\section{Results and discussions}
\label{}

\subsection{Structural properties}

\begin{figure}[htp]
\includegraphics[scale=0.4300,angle=270]{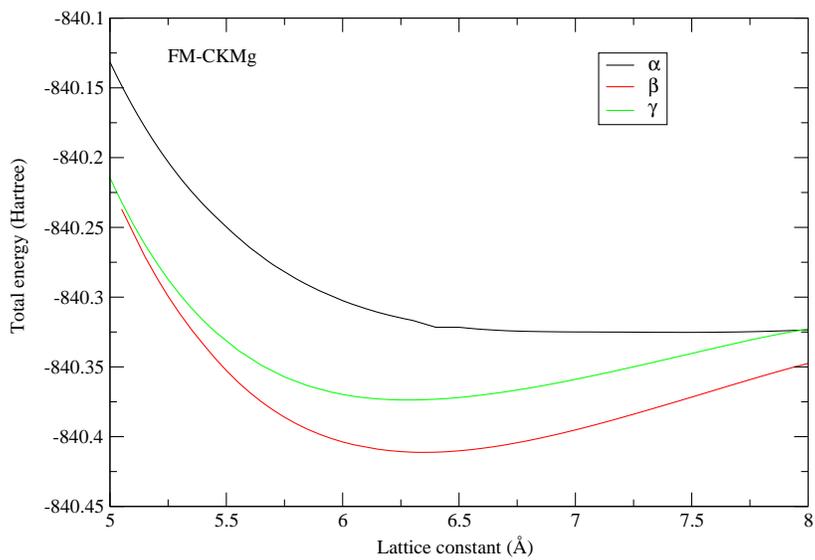}
\includegraphics[scale=0.4300,angle=270]{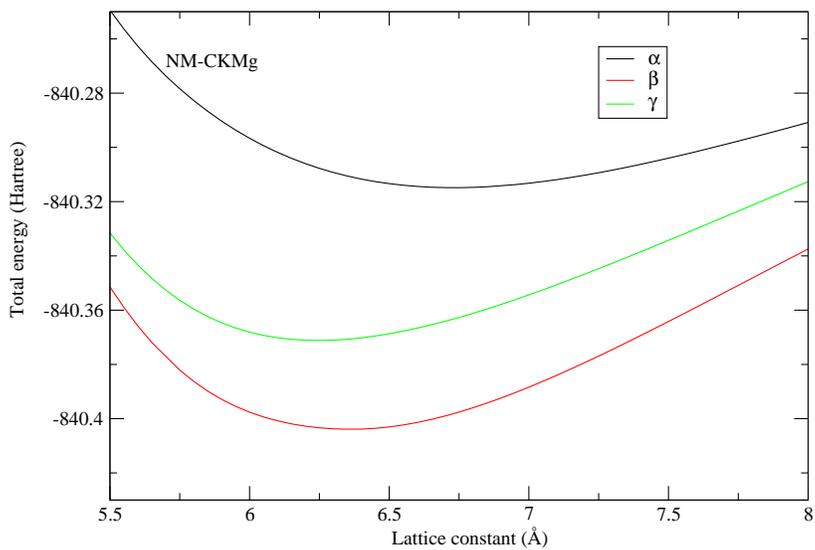}
\caption{The total energies per unit cell as a function of lattice constants for $\alpha$, $\beta$ and $\gamma$ structure for the CKMg in the FM phase and NM phase, respectively.}
\label{1}
\end{figure}

First, to get the equilibrium lattice constants of the half-Heusler alloys CKMg and SiKMg,
we calculate the total energy as a function of lattice constants for the three possible
atomic arrangements. Fig.\ref{1} shows the total energy curves with respect to the
lattice constants of $\alpha$, $\beta$, $\gamma$ structures in the ferromagnetic (FM) and the nonferromagnetic (NM) phases for CKMg. Fig.2 is for SiKMg. From
the two figures, we can get that for both alloys, the $\beta$       structure is the most stable among the three in both FM and NM phases. Then we compare the total energies of the two alloys in the FM
and NM phases of the $\beta$ structure. The total energy-lattice constant relation curves
are shown in Fig.\ref{3}. It implys that the FM state is more stable than the NM state. We get the lattice constant of equilibrium is 6.345 \AA\ for CKMg and 7.160 \AA\ for SiKMg by using
Murnaghan equation \cite{22}. The equilibrium lattice constants, bulk modulus, the first order derivative of the modulus on volume, cohesive energy and formation energy we calculated are displayed in Table 1.

Formation energy ($\Delta H_{for}^{XKMg}$) shows the stability of the compound with respect to decomposition into bulk constituents. For each alloy, the formation energy can be expressed by  
\begin{equation}
\Delta H_{for}^{XKMg} = E_{total}^{XKMg}-(E_{bulk}^{X}+E_{bulk}^{K}+E_{bulk}^{Mg}),
\end{equation}
Where $E_{bulk}^X$, $E_{bulk}^K$, and $E_{bulk}^{Mg}$ are the total energies per atom for bulk X (X=C, Si), K and Mg, respectively.
In general, a negative value of formation energy indicates compounds could be synthesized easily. Both the alloys are likely to be synthesized experimentally due to the negative formation energy (see Table.1).

The cohesive energies of the two alloys can be obtained from the equation: 

\begin{equation}
E_{coh} = (E_{atom}^{X}+E_{atom}^{K}+E_{atom}^{Mg})-E_{total}^{XKMg},
\end{equation} 
where $E_{total}^{XKMg}$ is the total energies of the considered compound, $E_{atom}^{X}$ (X=C, Si), $E_{atom}^{K}$ and $E_{atom}^{Mg}$ are the energies of isolated constituent atoms in each compound. In general, a positive value of cohesive energy indicates the stability of the material. Therefore we can confirm the alloys we study are excepted to be stable. 

\subsection{Electronic and magnetic properties}
Fig.\ref{4} displays the total density of states of CKMg and SiKMg alloys under equilibrium lattice constants. From the figure we can see, for the SiKMg alloy, the spin-up states show semiconductor character while the spin-down states show metallic nature. So we can get that the SiKMg alloy is a half-metallic ferromagnet. The spin magnetic moments are 1 $\mu_B$ for the two alloys per formula unit (See Table 1). There are 7 valence electrons in SiKMg alloy, and the total magnetic moment of 1 $\mu_B$ per formula unit complies with the Slater-Pauling behavior of HM half Heusler  alloys
\begin{equation}
M_{tot}=(8-Z_{tot}) \mu_B
\end{equation}
where $Z_{tot}$ is the number of total valence electrons and $M_{tot}$ the total magnetic moment per formula unit. We offer the band gap and spin-flip gap in Table 1. Fig.\ref{5} displays the band structure of the SiKMg alloy. The band gap is formed by the valence band maximum (VBM) and the  conduction band minimum (CBM) at $\Gamma$ point.

Now, we have a discussion on the CKMg alloy. From Fig.\ref{4}, we can see the density of states (DOS) in both the spin down and spin up band structures get through the Fermi level, so the half Heusler alloy is not a half-metal, but a nearly half-metal. According to the definition, the spin polarization(P) can be expressed by

\begin{equation}
P =\frac{ N\downarrow (\varepsilon _F )-N\uparrow (\varepsilon _F)}{N\downarrow (\varepsilon _F )+N\uparrow (\varepsilon _F)},
\end{equation}
where $N\downarrow (\varepsilon _F)$ and $N\uparrow (\varepsilon _F)$ are the DOS of majority-spin and minority-spin at Fermi level, respectively. The values of $N\downarrow (\varepsilon _F)$ and $N\uparrow (\varepsilon _F)$ can be obtained from Fig.\ref{4}. Through Eq.(4), we can calculate out that the spin polarization of the SiKMg alloy is 99.99 \%\ under its equlibrium lattice constant. The spin polarization is so high that we can get the half-metallicity of the CKMg by applying a very low pressure, which will be discussed in section 3.4.

\subsection{Stability of half-metallicity }
HM materials are usually used in spintronic devices in the form of thin films or multilayers, the lattice constant will have a change when the films or multilayers are grown on appropriate substrate, and the corresponding half-metallicity may be destroyed. So it is meaningful to study the robustness of the half-metallicity with respect to variation of the interatomic distance. Fig.\ref{6} gives the relation between the lattice constant and external pressure for the SiKMg alloy. Fig.\ref{7} shows the change of total magnetic moment as a function of the lattice constants and pressures. As mentioned previously, integer magnetic moments are an unique property of half-metallic. The total magnetic moment remains to be an integer with the lattice constant changes in the ranges of  5.90 \AA\ to 7.16 \AA.\  Namely SiKMg alloy can preserve half-metallic character when its lattice constant is changed by 17.6 \%.\  From Fig.\ref{6} we can get that the critical lattice constant with half-metallic character is equivalent to add the external pressure of 25.22 GPa. Therefore, the SiKMg alloy has stable half-metallic character, and it could be applied in the thin films and layers for spin valves and magnetic tunnel junctions and can be used as electrodes in tunnel junctions.

\subsection{Effect of pressure}

From Fig.\ref{7}, we can see for the SiKMg alloy, the magnetic moment keeps 1.00 $\mu_B$  when the pressure is in the range from 0.00 GPa to about 25.22 GPa.  When the external pressure is up to 35.99 GPa, the magnetic moment becomes zero, which indicates the transition from FM to NM state.

Fig.\ref{8} displays the density of states with different pressures of the SiKMg alloy. From the figure we can get that the $p$ states of Si atoms make the main contribution to the density of states from -2.7 to 0 eV below the Fermi level, indicating the magnetic moment mainly carried by Si atoms, and the total states from 4 to 14 eV are mainly contributed by the $d$ states of K atoms. A $p$-$d$ hybridization occurs around the Fermi level. The Fermi level slowly shifts to the edge of VBM and finally passes through it, which means the changing from half-metallic to metallic characters. 

 As mentioned in section 3.2, the CKMg alloy has such a very high spin polarization that just by applying a very low pressure it may transform into a half-metal. In the following we will discuss the effect of pressure on the CKMg alloy. Fig.\ref{9} and Fig.\ref{10} show the changes of magnetic moment and band gap under pressure. From Fig.\ref{9}, we can see, the magnetic moment  keeps 1 $\mu_B$ when pressure in the range from 0 GPa to 20.35 GPa. The magnetic moment becomes zero when the external pressure beyond 23.60 GPa. From Fig.\ref{10} we can get that CKMg will have a band gap just by applying a pressure of 0.23 GPa. This shows CKMg is a good spintronic functional material. In the range from 0.23 GPa to 20.35 GPa, CKMg keeps having band gaps, while upon 20.35 GPa, the gaps vanish. On the whole, CKMg alloy keeps being HM when pressure in the range from 0.23 GPa to 20.35 GPa, a ferromagnet from 20.35 GPa to 23.60 GPa and a paramagnet beyond 23.60 GPa.

\section{Conclusions}
\label{}
In this paper, we study the structural, electronic, and magnetic properties of half-Heusler alloys CKMg and SiKMg. Our results show the SiKMg alloy is a half-metallic ferromagnet. The half-metallicity can be preserved up to 17.6 \%\ compression of lattice constant with respect to its equilibrium lattice constant. CKMg is a nearly half-metal with a very high spin-polarization of 99.99 \%\ under its equilibrium lattice constant and it will transform into a half-metal when a low pressure of 0.23 GPa applied, showing CKMg is a very promising functional material. So our results will offer some valuable hints to spintronic material design and development.





\section*{Reference}

\bibliographystyle{elsarticle-num} 
\bibliography{all}

\begin{thebibliography}{10}
\expandafter\ifx\csname url\endcsname\relax
  \def\url#1{\texttt{#1}}\fi
\expandafter\ifx\csname urlprefix\endcsname\relax\def\urlprefix{URL }\fi
\expandafter\ifx\csname href\endcsname\relax
  \def\href#1#2{#2} \def\path#1{#1}\fi

\bibitem{1}
R.~De~Groot, F.~Mueller, P.~Van~Engen, K.~Buschow, New class of materials:
  half-metallic ferromagnets, Physical Review Letters 50~(25) (1983) 2024.

\bibitem{2}
F.~Jedema, A.~Filip, B.~Van~Wees, Electrical spin injection and accumulation at
  room temperature in an all-metal mesoscopic spin valve, Nature 410~(6826)
  (2001) 345--348.

\bibitem{3}
W.~Song, E.~Zhao, J.~Meng, Z.~Wu, et~al., Near compensated half-metal in
  sr2nioso6, Journal of Chemical Physics 130~(11) (2009) 114707.

\bibitem{4}
J.~Li, Y.~Li, X.~Dai, X.~Xu, Band structure calculations for heusler phase co 2
  ybi and half-heusler phase coybi (y= mn, cr), Journal of Magnetism and
  Magnetic Materials 321~(5) (2009) 365--372.

\bibitem{5}
S.~Soeya, J.~Hayakawa, H.~Takahashi, K.~Ito, C.~Yamamoto, A.~Kida, H.~Asano,
  M.~Matsui, Development of half-metallic ultrathin fe 3 o 4 films for
  spin-transport devices, Applied physics letters 80~(5) (2002) 823--825.

\bibitem{6}
V.~Ko, G.~Han, Y.~Feng, Electronic band structure matching for half-and
  full-heusler alloys, Journal of Magnetism and Magnetic Materials 322~(20)
  (2010) 2989--2993.

\bibitem{7}
G.~Liu, X.~Dai, H.~Liu, J.~Chen, Y.~Li, G.~Xiao, G.~Wu, Mn 2 co z (z= al, ga,
  in, si, ge, sn, sb) compounds: Structural, electronic, and magnetic
  properties, Physical Review B 77~(1) (2008) 014424.

\bibitem{8}
T.~Ogawa, M.~Shirai, N.~Suzuki, I.~Kitagawa, First-principles calculations of
  electronic structures of diluted magnetic semiconductors (ga, mn) as, Journal
  of magnetism and magnetic materials 196 (1999) 428--429.

\bibitem{9}
S.~Chatterjee, V.~Singh, A.~Deb, S.~Giri, S.~De, I.~Dasgupta, S.~Majumdar,
  Magnetic properties of ni2+ xmn1-xin heusler alloys: Theory and experiment,
  Journal of Magnetism and Magnetic Materials 322~(1) (2010) 102--107.

\bibitem{10}
V.~Alijani, J.~Winterlik, G.~H. Fecher, S.~S. Naghavi, C.~Felser, Quaternary
  half-metallic heusler ferromagnets for spintronics applications, Physical
  Review B 83~(18) (2011) 184428.

\bibitem{11}
H.~Luo, G.~Liu, F.~Meng, L.~Wang, E.~Liu, G.~Wu, X.~Zhu, C.~Jiang,
  Slater--pauling behavior and half-metallicity in heusler alloys mn 2 cuz (z=
  ge and sb), Computational Materials Science 50~(11) (2011) 3119--3122.

\bibitem{12}
G.~Jaiganesh, R.~Eithiraj, G.~Kalpana, Theoretical study of electronic,
  magnetic and structural properties of mo and w based group v (n, p, as, sb
  and bi) compounds, Computational Materials Science 49~(1) (2010) 112--120.

\bibitem{13}
I.~Galanakis, P.~Mavropoulos, Zinc-blende compounds of transition elements with
  n, p, as, sb, s, se, and te as half-metallic systems, Physical Review B
  67~(10) (2003) 104417.

\bibitem{14}
B.-G. Liu, Robust half-metallic ferromagnetism in zinc-blende crsb, Physical
  Review B 67~(17) (2003) 172411.

\bibitem{15}
G.~Gao, K.~Yao, E.~{\c{S}}a{\c{s}}{\i}o{\u{g}}lu, L.~Sandratskii, Z.~Liu,
  J.~Jiang, Half-metallic ferromagnetism in zinc-blende cac, src, and bac from
  first principles, Physical Review B 75~(17) (2007) 174442.

\bibitem{16}
S.~Dong, H.~Zhao, First-principles studies on magnetic properties of rocksalt
  structure mc (m= ca, sr, and ba) under pressure, Applied Physics Letters
  98~(18) (2011) 182501.

\bibitem{17}
F.~Ahmadian, R.~Alinajimi, First-principles study of half-metallic properties
  for the heusler alloys sc 2 crz (z= c, si, ge, sn), Computational Materials
  Science 79 (2013) 345--351.

\bibitem{18}
K.~Koepernik, H.~Eschrig, Full-potential nonorthogonal local-orbital
  minimum-basis band-structure scheme, Physical Review B 59~(3) (1999) 1743.

\bibitem{19}
K.~Koepernik, B.~Velick{\`y}, R.~Hayn, H.~Eschrig, Analytic properties and
  accuracy of the generalized blackman-esterling-berk coherent-potential
  approximation, Physical Review B 58~(11) (1998) 6944.

\bibitem{20}
J.~P. Perdew, K.~Burke, M.~Ernzerhof, Generalized gradient approximation made
  simple, Physical review letters 77~(18) (1996) 3865.

\bibitem{21}
F.~D. Murnaghan, Finite deformation of an elastic solid.

\bibitem{22}
J.~Liu, R.~Zhan, L.~Li, H.-N. Dong, Condensed matter: Electronic structure,
  electrical, magnetic, and optical properties: Magnetic properties of several
  potential rocksalt half-metallic ferromagnets based on the first-principles
  calculations, Chinese Physics B 20~(7) (2011) 077101.

\end{thebibliography}


\begin{figure}[htp]
\centering
\includegraphics[scale=0.43, angle=270]{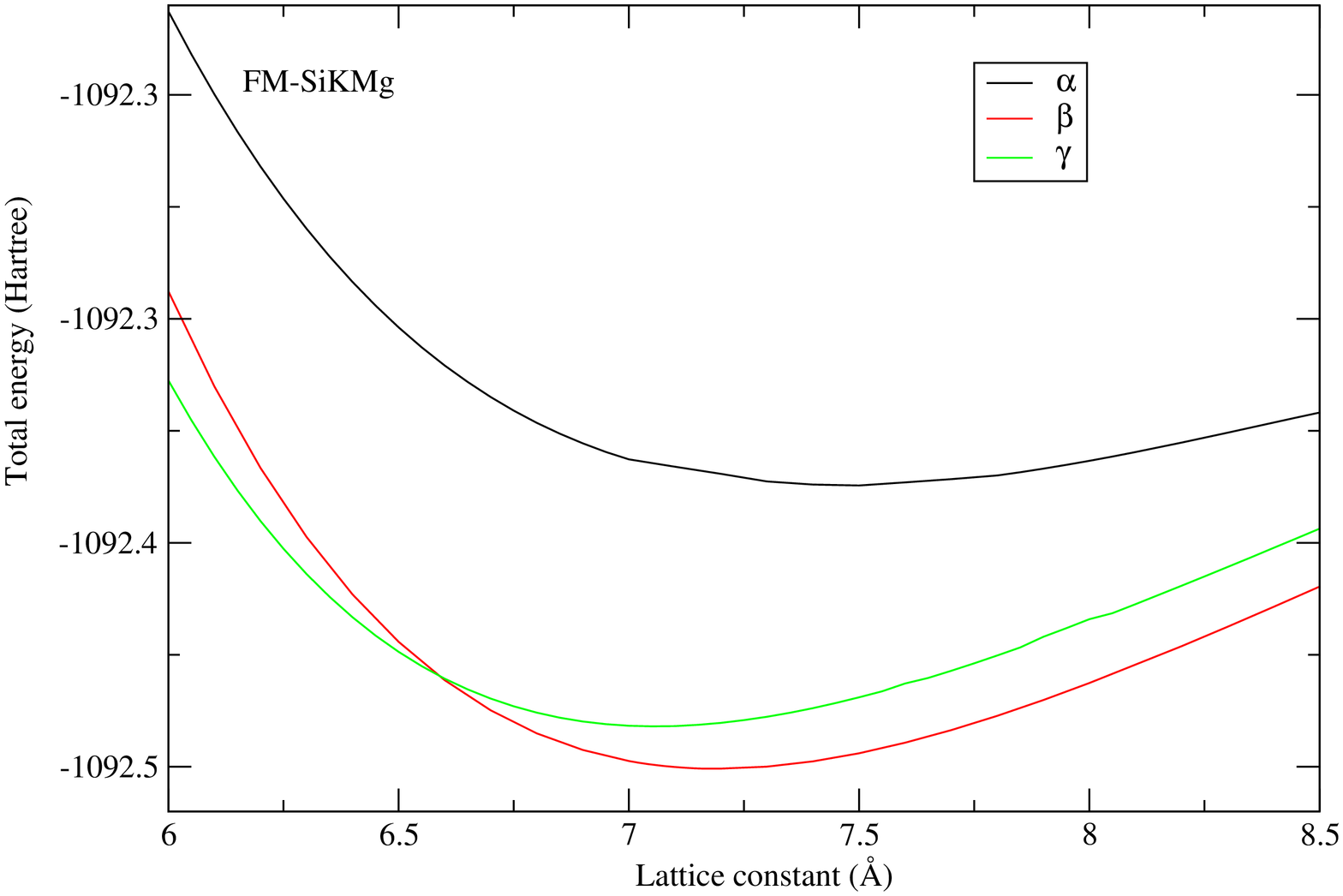}
\includegraphics[scale=0.43, angle=270]{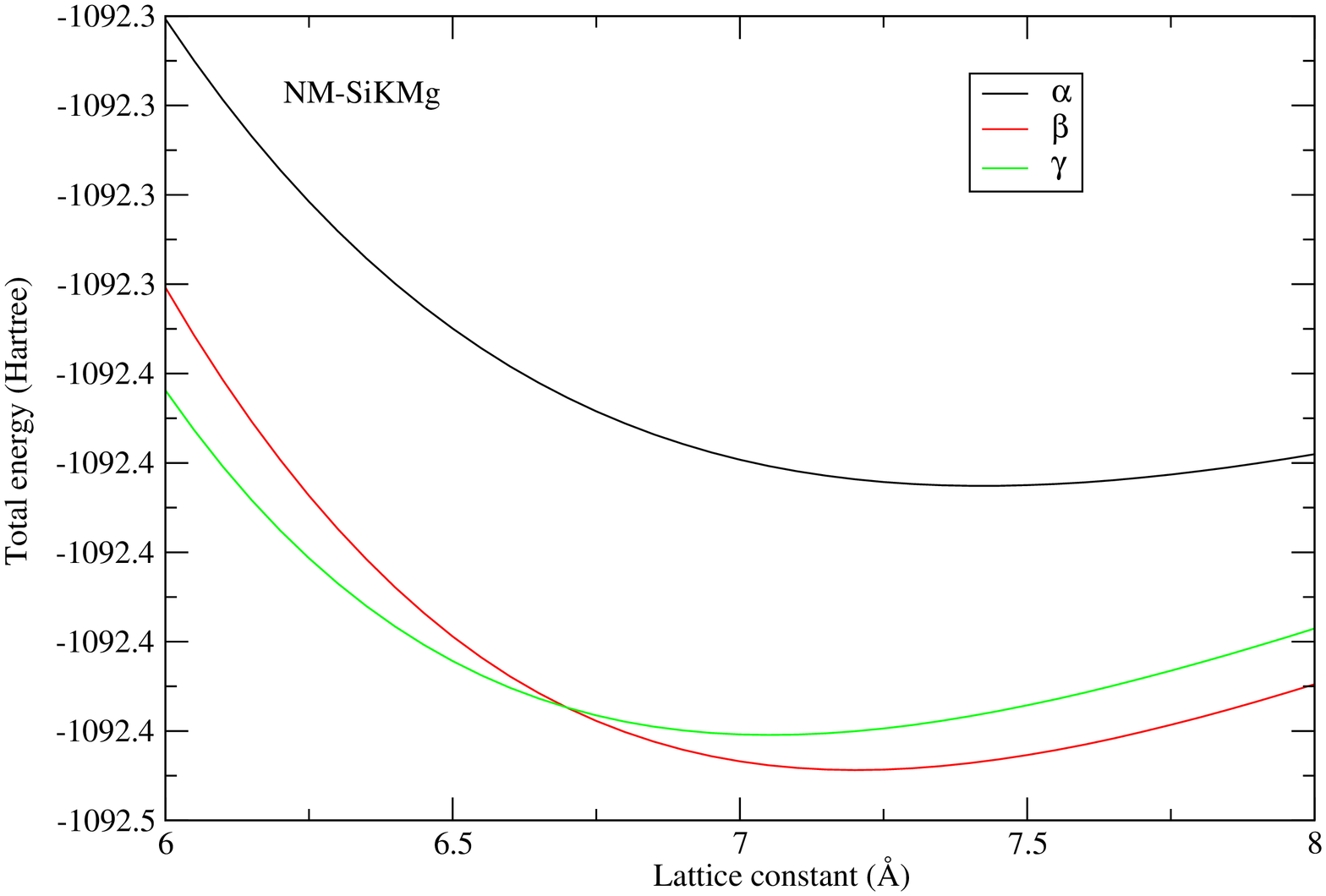}
\caption{The total energies per unit cell as a function of lattice constants for $\alpha$, $\beta$ and $\gamma$ structure for the SiKMg in the FM phase and NM phase, respectively.}
\label{2}
\end{figure}
\begin{figure}[htp]
\centering
\includegraphics[scale=0.600,angle=270]{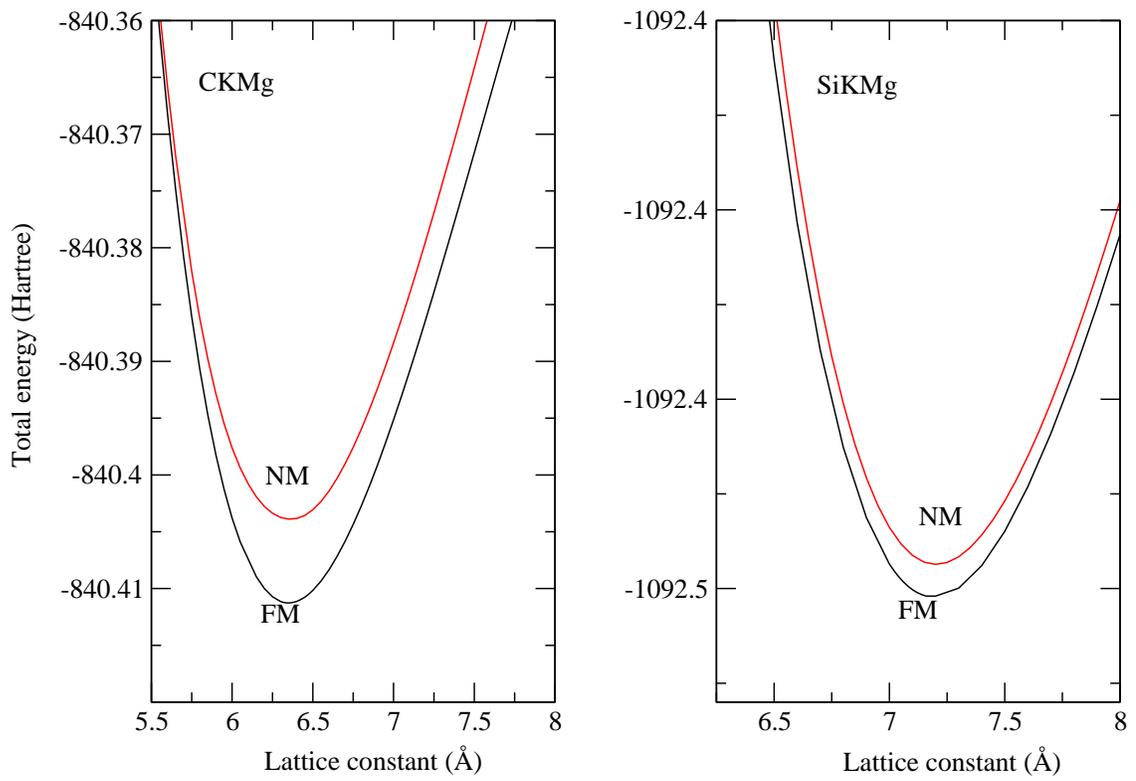}
\caption{The $\beta$-structure total energies of nonmagnetic (NM) and ferromagnetic phase (FM) as a function of the lattice constant for CKMg and SiKMg.}
\label{3}
\end{figure}

\begin{figure}[htp]
\centering
\includegraphics[scale=0.4800,angle=270]{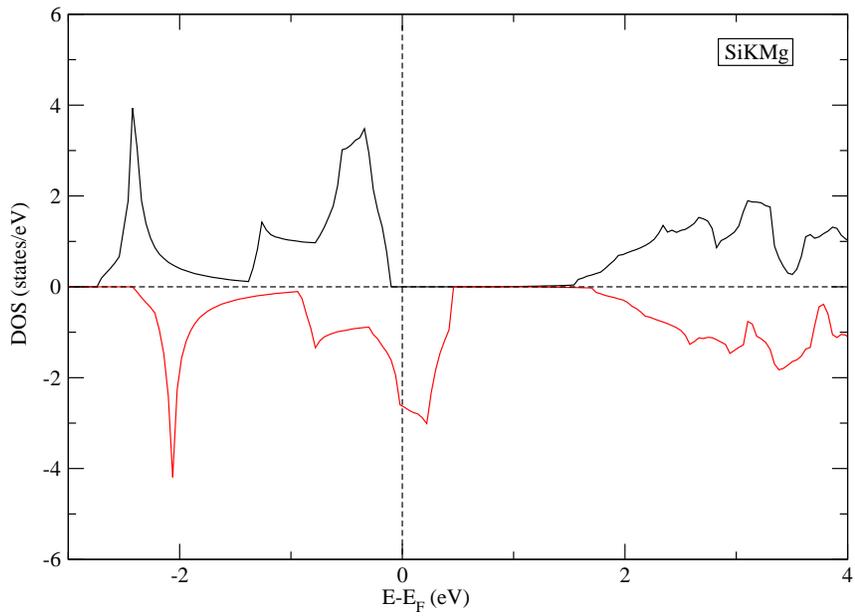}
\includegraphics[scale=0.4800,angle=270]{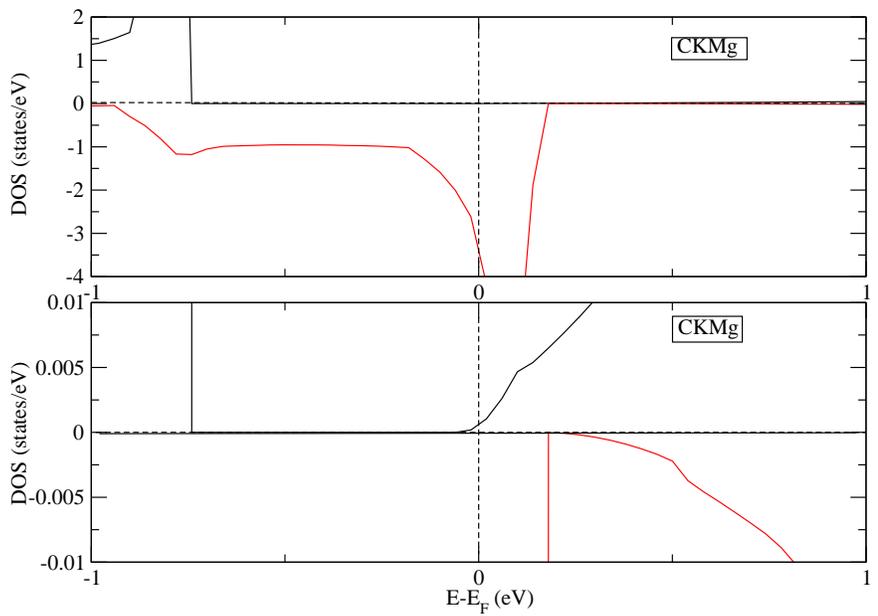}
\caption{The total DOS for SiKMg and amplified total DOS near the Fermi level for CKMg.}
\label{4}
\end{figure}

\begin{figure}[htp]
\centering
\includegraphics[scale=0.550,angle=270]{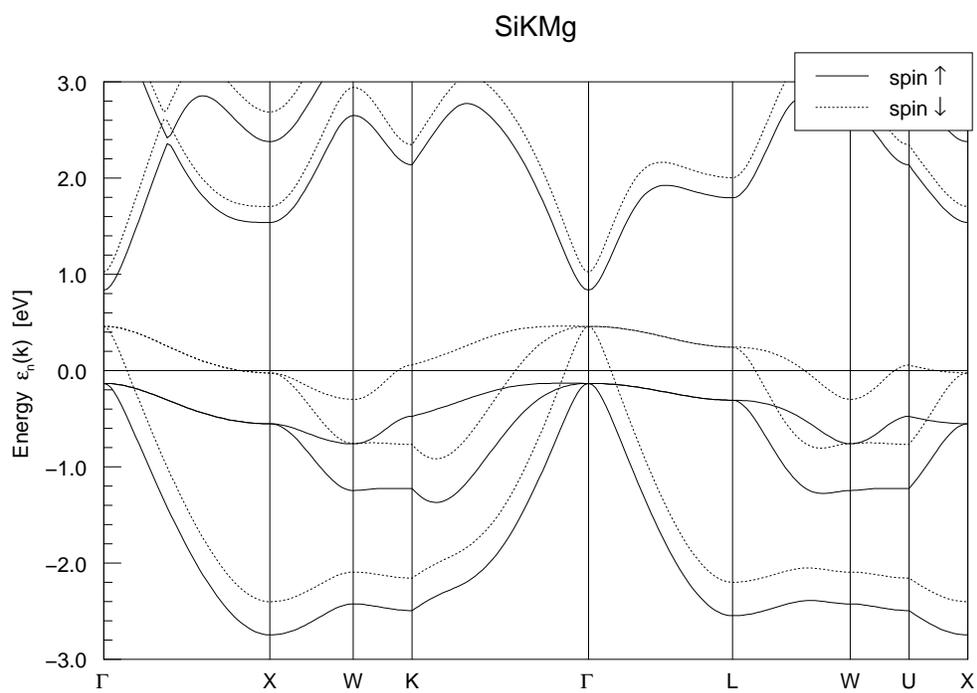}
\caption{Band structure for SiKMg at equilibrium lattice constant. }
\label{5}
\end{figure}

\begin{figure}[htp]
\centering
\includegraphics[scale=1]{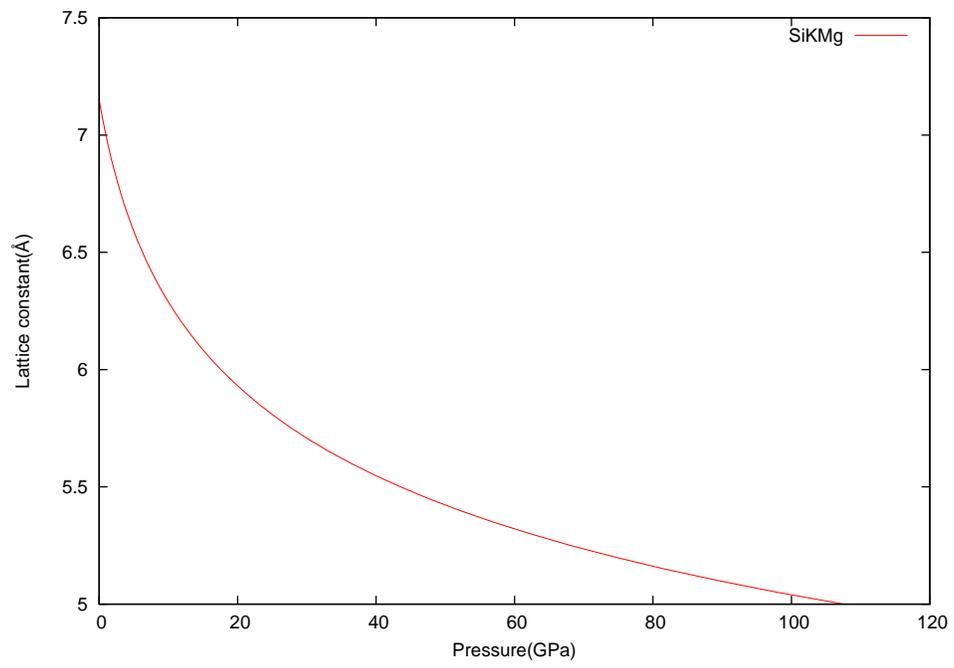}
\caption{Pressure-lattice constant relation for SiKMg.}
\label{6}
\end{figure}

\begin{figure}[htp]
\centering
\includegraphics[scale=0.600,angle=270]{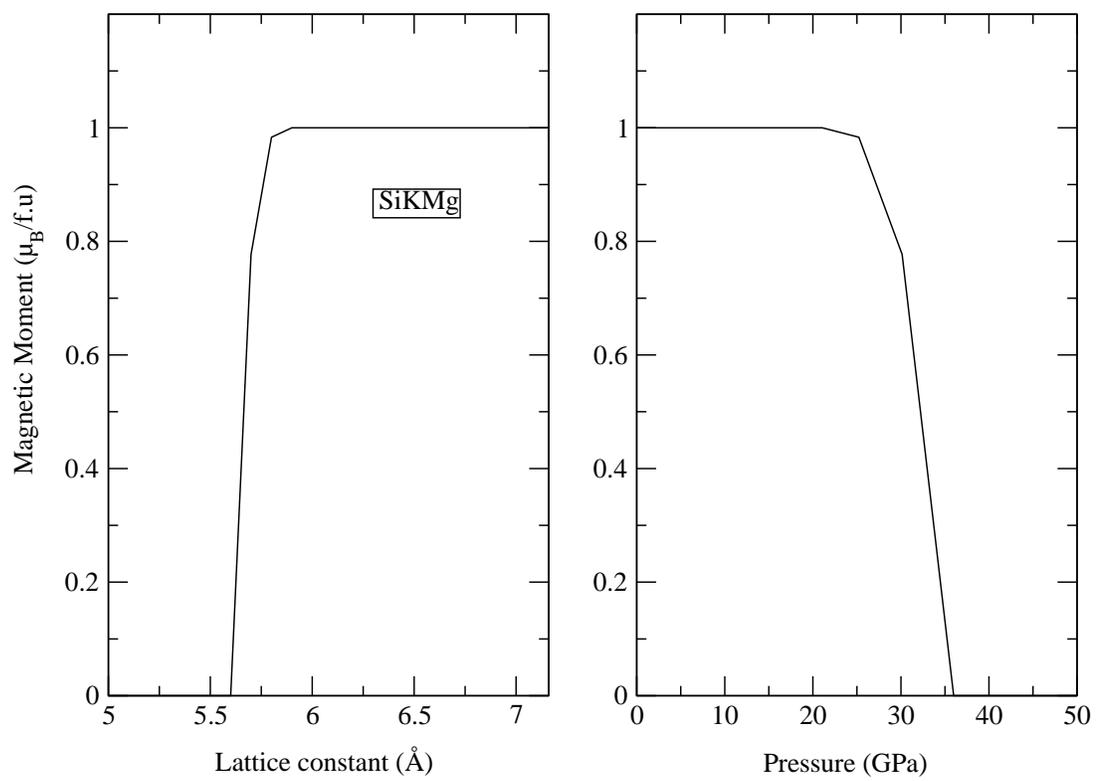}
\caption{The magnetic moment as a function of the lattice constant and pressure for SiKMg alloy.}
\label{7}
\end{figure}

\begin{figure}[htp]
\centering
\includegraphics[scale=0.5000,angle=270]{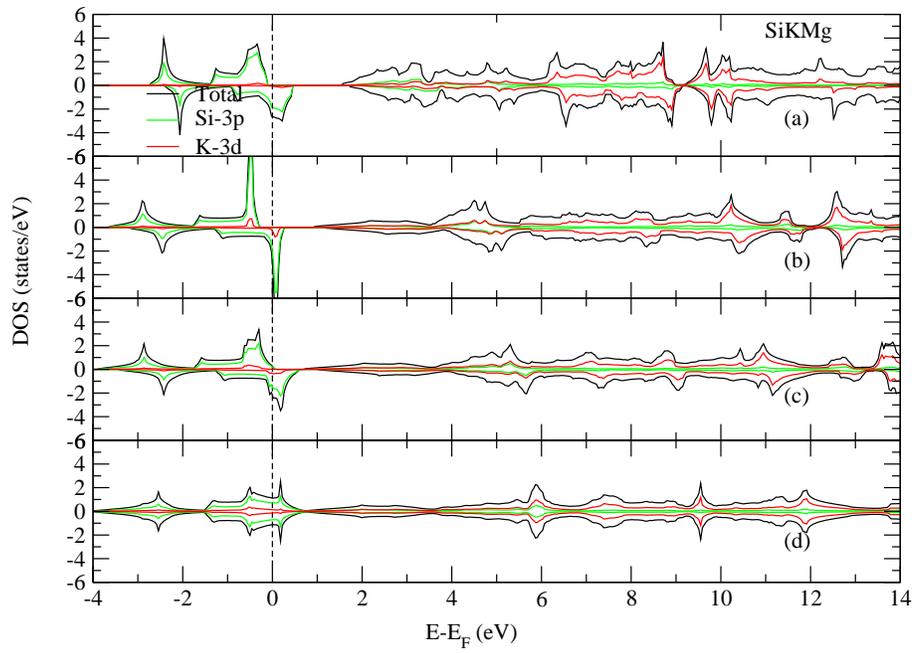}
\caption{Spin resolved electronic density of states of SiKMg under pressures: (a) 0.00 GPa, (b) 14.56 GPa, (c) 30.14 GPa, (d) 42.95 GPa.     }
\label{8}
\end{figure}

\begin{figure}[htp]
\centering
\includegraphics[scale=0.500,angle=270]{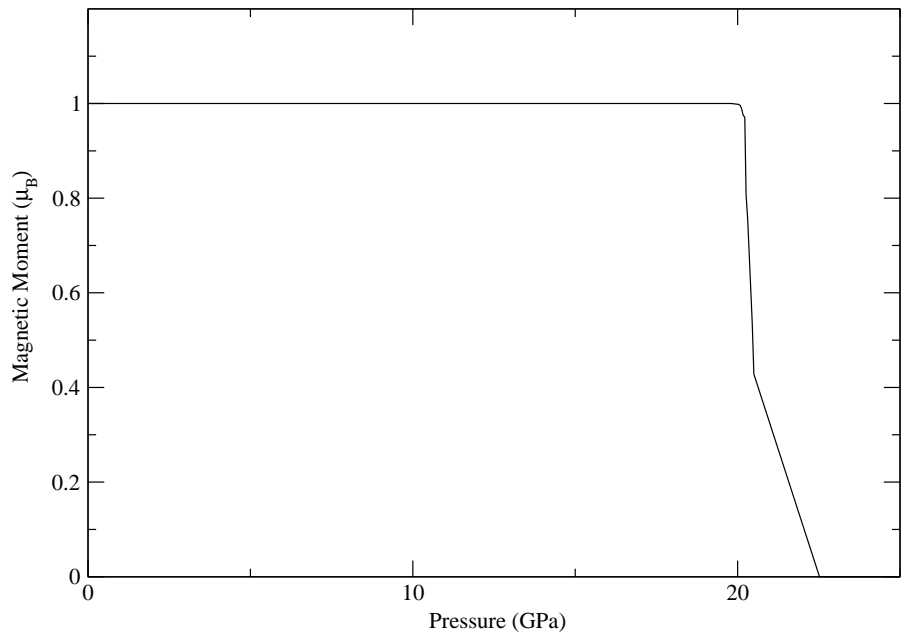}
\caption{The changes of magnetic moment for CKMg under pressure.}
\label{9}
\end{figure}

\begin{figure}[htp]
\centering
\includegraphics[scale=0.500,angle=270]{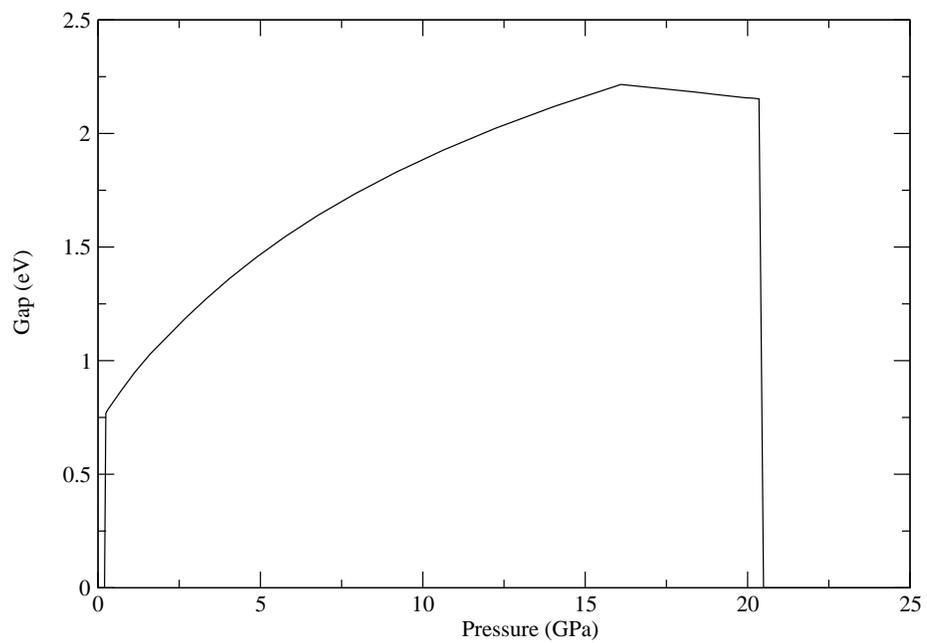}
\caption{The changes of band gap for CKMg under pressure.}
\label{10}
\end{figure}

 \begin{table}[b]
\caption{\label{tab:test}Calculated equilibrium lattice ($a_0$), bulk modulus $B$, the first order derivative of the modulus on volume ($B^{'}$), cohesive energy ($E_{coh}$), formation energy per unit cell ($\Delta H$), total spin magnetic moment ($M_t$), band gap ($E_{bg}$), and spin-flip gap ($E_{sfg}$) are shown for SiKMg and CKMg, respectively.  }
\begin{tabular}{cccccccccccccccc}
\toprule
XKMg& $a_0$(\AA) & B(GPa)&$B^{'}$ &$E_{coh}(eV)$&$\Delta H$(eV)&$M_t$($\mu_B$)&$E_{bg}$(eV)&$E_{sfg}$(eV)\\
\midrule
CKMg&6.345&13.3&4.12&6.02&-5.85&1.00&0.000&0.000\\
SiKMg&7.160&13.6&2.96&5.60&-5.41&1.00&0.949&0.105\\
\bottomrule

	\end{tabular}
\end{table}

\end{document}